\documentclass[aps,prl,twocolumn,showkeys]{revtex4-1}

\usepackage{graphicx} 
\usepackage{amsfonts,amsmath,dsfont} 
\usepackage{mathrsfs} 
\usepackage{setspace}
\usepackage[utf8]{inputenc} 
\usepackage{url}
\usepackage{comment} 

\DeclareMathAlphabet{\mathbbold}{U}{bbold}{m}{n}

\DeclareMathOperator{\Tr}{Tr}

\begin{document}
\title{On the Existence of Monodromies for the Rabi model}

\author{Bruno Carneiro da Cunha}
\email{bcunha@df.ufpe.br}
\affiliation{Departamento de Física, Universidade Federal de Pernambuco,
50670-901, Recife, Pernambuco, Brazil} 
\author{Manuela Carvalho de Almeida}
\email{mcalmeida.13@gmail.com}
\author{Amílcar Rabelo de Queiroz}
\email{amilcar@unb.br}
\affiliation{Instituto de Física,
Universidade de Bras\'ilia, Caixa Postal 04455, 70919-970, Brasília,
DF,Brazil}

\begin{abstract}
We discuss the existence of monodromies associated with the singular
points of the eigenvalue problem for the Rabi model. The complete
control of the full monodromy data requires the taming of the Stokes
phenomenon associated with the unique irregular singular point. The
monodromy data, in particular the composite monodromy, are written in
terms of the parameters of the model via the isomonodromy method and
the tau-function of the Painlev\'e V. These data provide a systematic
way to obtain the quantized spectrum of the Rabi model.   
\end{abstract}


\date{\today}

\maketitle

\section{Introduction}

The Rabi model \cite{PhysRev.49.324,PhysRev.51.652} describes the
interaction of a single 
electromagnetic mode -- a simple harmonic oscillator -- with matter --
a two-level quantum system. It is a quite simple model yet it has an
interesting and rich spectrum. Recently, it has attracted vividly
attention due to experimental and mathematical reasons. From applied
and experimental physics side, there have emerged interesting applications
in quantum optics and quantum computation because of good prospects of
experimental realization using Josephson junctions, traped ions and
others (see references in \cite{braak-PhysRevLett.107.100401}). From the mathematical side, for a long time it has been a
challenge to prove its exactly solvability. And so, finally, as
recently as 2011, Braak \cite{braak-PhysRevLett.107.100401} solved the model in a Bargmann
(coherent state) representation obtaining in a systematic way its
regular spectrum as zeroes of a transcendental function. The
exceptional part of the spectrum were already known since the late
1970's \cite{Judd-1979}, but it can also be obtained in the framework proposed by
Braak \cite{Zhong2013}. 

This fact has opened up a whole new set of interesting problems in
mathematical physics. One issue regards the notion of integrability of
the Rabi model. At face value, the Rabi model could be a first
instance of an exactly solvable yet not integrable model in
mathematical physics. A controversy ensues since the proper definition
of integrability in quantum physics -- or even in mathematics -- seems
not to be clear cut \cite{Caux-Mossel,Batchelor2014}. Braak claimed that the Rabi model is
integrable in a new quantum integrability criterion coined by
himself. A system is dubbed Braak integrable if there are $f=f_c+f_d$
quantum numbers classifying the eigenstates uniquely, where $f_c,f_d$
stand for the number of continuous and discrete degrees of freedom
(d.o.f.), respectively. For the Rabi model, this number is
$f_c=f_d=1$, one harmonic oscillator and one two-level system, so that
$f=2$.  These two quantum numbers arises from the $\mathds{Z}_2$ parity
symmetry of the Hamiltonian. Even in the lack of a better mathematical
formulation of this criterion -- still a pending project -- it served
as the guiding principle to the solution of this outstanding problem. 

Controversies aside, it would be desirable to frame this discussion in
a more conservative setup. Therefore, Batchelor and Zhou \cite{Batchelor2014} raised the
issue of whether the Rabi model is Yang-Baxter integrable (YBI). They
succeed to show YBI for two special points in the space of parameters
of the Rabi model. For a generic point in this moduli, YBI is still an
open question. 

Another important issue arising from the Braak's work is that of
finding an universal method that applies to a wide variety of models
involving coupling of a boson mode with a two-level system in the
Bargmann representation. Such a program has been developed by
Maciejewski et al. \cite{Maciejewski2012,Maciejewski2014}. They have
used a method framed in terms of the Wronskians, that is, a $2\times
2$ matrix containing both the wave-function and its derivative in the 
neighbourhood of a singular point.   

The present work advances some enlightenment in the direction of
proving YBI for a generic point in the moduli of parameters of the
Rabi model. The departing point is the Bargmann representation of the
Rabi model. In this representation, it can be easily shown \cite{Zhong2013}
that the Rabi model is described by a confluent Heun equation. Here we use
the known fact \cite{Jimbo:1982,Iwasaki:1991} that the monodromy data
of these equations can be cast in terms of Painlevé V transcendent
tau-function \cite{Jimbo:1981b} via isomonodromy equations. The global
properties, relevant for the problem of the spectrum and for the YBI,
are encoded in the notion of composite monodromies. We then present
the composite monodromy parameter of the Rabi model. Finally we
discuss in general terms how one could use this composite monodromy
parameter to obtain the Rabi model's spectrum.   

The novelty of our work consists in the presentation of the
monodromies associated with the singular points of the ODE arising
from the eigenvalue problem of the Rabi model. We then discuss the
relevance of the Stokes phenomenon in order to have a complete monodromy
data set. We conjecture that it is the emergence of the Stokes phenomenon and the
need of extra parameters in the monodromy data set that rendered extra
difficulties in the full demonstration of the Yang-Baxter integrability of the Rabi model. 

This work is organized as follows: we first write the Rabi model as a
standard Fuchsian system, and then discuss the mododromies around the
singular points. A special situation happens for the monodromy around
the unique irregular singular point, the point at infinity, giving rise to the Stokes phenomenon,
which we discuss in detail. As an outcome we obtain the general group
relation for the monodromies and how it is related with Yang-Baxter
equations. We next discuss the isomonodromy method aiming at writing
the composite monodromy parameter in terms of the monodromy parameter
at the irregular point and the stokes parameters. On the other hand,
the existence of monodromy matrices for our original system are
obtained from the tau-function of the Painlevé V. We finally obtain
the composite monodromy parameter in terms of the parameters of the
Rabi model which provides in a systematic way the quantized spectrum of the
Rabi model.

\section{Rabi and its Monodromies}

The quantum Rabi model (QRM) is described by the Hamiltonian
\begin{equation}
  H_R=a^\dagger a + \Delta \sigma^z + g\sigma^x\left(a^\dagger + a \right),
\end{equation}
where the boson mode is described by $[a,a^\dagger]=1$, the fermion
mode by the Pauli matrices, $\Delta$ is the level separation of the
fermion mode and $g$ is the boson-fermion coupling.   

The QRM can be written in terms of two copies of Jaynes-Cummings model
(JCM), each with its appropriate chirality. Indeed, the chiral JCM
Hamiltonian reads 
\begin{equation}
  H_{JC}=a^\dagger a+\Delta \sigma^z + g \left( \sigma^+ a + \sigma^-
    a^\dagger\right), 
\end{equation}
and the anti-chiral one reads
\begin{equation}
  H_{\overline{JC}}=a^\dagger a+ \Delta \sigma^z + g \left( \sigma^- a
    + \sigma^+ a^\dagger \right), 
\end{equation}
so that
\begin{equation}
  H_R = \frac{1}{2}\left(~H_{JC} + H_{\overline{JC}}\right)=a^\dagger
  a + \Delta \sigma^z + g\sigma^x(a^\dagger + a), 
\end{equation}
where we have used $\sigma^x=(\sigma^++\sigma^-)/2$. It is important
to notice that  $[H_{JC},H_{\overline{JC}}]\neq 0$. 

Consider the Ansatz \cite{braak-PhysRevLett.107.100401,Zhong2013}
\begin{equation}
\label{ansatz-rabi}
  |\psi(a^\dagger)\rangle = f_1(a^\dagger) |0\rangle |+\rangle +
  f_2(a^\dagger) |0\rangle |-\rangle,  
\end{equation}
where the harmonic oscillator ground state is defined by $a|0\rangle =
0$, $\sigma^z|\pm\rangle = \pm |\pm\rangle$, and $f_i$, $i=1,2$, are
analytic functions of $a^\dagger$. We can now use Bargmann's
prescription 
\begin{equation}
  a^\dagger \mapsto w, \qquad a\mapsto \partial_w,
\end{equation}
so that $[a,a^\dagger]f(w)=f(w)$. Substituting the Ansatz into the
stationary Schroedinger equation, or the eigenvalue equation,
$H_R|\psi\rangle = E|\psi\rangle$, we obtain, after setting
$f_\pm=f_1\pm f_2$, 
\begin{equation}
\begin{gathered}
\partial_wf_+=\frac{E-gw}{w+g}f_+-\frac{\Delta}{w+g}f_- \\
\partial_wf_-=-\frac{\Delta}{w-g}f_++\frac{E+gw}{w-g}f_-.
\end{gathered}
\end{equation}
Eliminating say $f_+$ in terms of $f_-$, this results in a second
order linear differential equation for $f_+$ (and $f_-$) which could
be brought to a confluent Heun equation \cite{Zhong2013}. Let us define 
\begin{equation}
z=-2g(w+g),\quad\quad
\Phi(z)=\begin{pmatrix}
f^{(1)}_+ & f^{(2)}_+\\
f^{(1)}_- & f^{(2)}_-
\end{pmatrix},
\end{equation}
where $f^{(1,2)}_\pm$ are the two linearly independent solutions of
the system above. The fundamental matrix $\Phi(z)$ is then
invertible and unique up to right multiplication of a constant
matrix. With this change of variables, we can bring the model to a 
standard Fuchsian form:  
\begin{equation}
\frac{d\Phi}{dz}\Phi^{-1}=\frac{1}{2}\sigma_3+\frac{1}{z}A_0+\frac{1}{z-t}A_t,
\label{eq:schlesinger}
\end{equation}
with $t=-4g^2$ and
\begin{equation}
\begin{gathered}
\sigma_3=
\begin{pmatrix}
1 & 0 \\
0 & -1
\end{pmatrix},\quad
A_0=
\begin{pmatrix}
E+g^2 & -\Delta \\
0 & 0 
\end{pmatrix},
\\[0.2cm]
A_t=
\begin{pmatrix}
0 & 0 \\
-\Delta & E+g^2
\end{pmatrix}.
\end{gathered}
\end{equation}

The system \eqref{eq:schlesinger} has two regular singular points at
$z_i=0,t$, $i=0,t$ and an irregular singular point at
$z_\infty=\infty$ with Poincaré index 1. The analytical structure of
the system near the regular singular point is characterized by the
monodromy matrices $M_0$ and $M_t$, defined as the effect of an
analytical continuation around the corresponding singular point,
\begin{equation}
\Phi((z-z_i)e^{2\pi i}+z_i)=\Phi(z)M_i, \qquad i=0,t.
\end{equation}
One notes that, since any two set of solutions $\Phi(z)$ are related
by right multiplication, the monodromy matrices are defined up to an
overall conjugation. Moreover, one can choose initial conditions
of \eqref{eq:schlesinger} so that, near a regular singular point
$z_i$, one has 
\begin{equation}
\left.\Phi(z)\right|_{z\approx z_i}=\begin{pmatrix}
(z-z_i)^{\alpha^+_i} & 0 \\
0 & (z-z_i)^{\alpha^-_i} \\
\end{pmatrix}.
\end{equation}
Therefore, one can see that generically the monodromy matrix $M_i$ can 
be written as 
\begin{equation}
M_i=g_i\begin{pmatrix}
e^{2\pi i\alpha^+_i} & 0 \\
0 & e^{2\pi i\alpha^-_i}
\end{pmatrix}
g_i^{-1},
\end{equation}
where $g_i\in {\rm SL}(2,\mathbb{C})$ are called the connection
matrices. They are also defined up to left multiplication. One
also notes that, for algebraic purposes, only the difference
$\theta_i=\frac{1}{2}(\alpha^+_i-\alpha_i^-)$ is important. Overall
shift of the coefficients $\alpha^{\pm}_i$ can be obtained by an
s-transformation of the solutions: $f_\pm(z)\rightarrow
(z-z_i)^{a}f_\pm(z)$. 

The system at the irregular singular point at $z=\infty$ is slightly
more complicated, due to the {\it Stokes phenomenon}. In order to
describe it, let us start by noting that close to $z=\infty$ the
solutions are of the form $f_\pm^{(1,2)}\approx e^{\pm z/2}$, and the
Frobenius series obtained at this point is only formal: its 
convergence radius is zero. One can see that generically near
infinity the system has the form
\begin{equation}
\frac{d\Phi}{dz}\Phi^{-1}=-\frac{1}{2}\sigma_3+\frac{A_0+A_t}{z}+{\cal O}(z^{-2}).
\end{equation} 
From the coefficient $A_\infty=-(A_0+A_t)$ of the $z^{-1}$ term one
can define the naive monodromy at $z=\infty$, given by the difference
of the eigenvalues
$\theta_\infty=\alpha^-_\infty-\alpha^+_\infty$. The epithet ``naive''
comes in because the monodromy structure around $z=\infty$ also
depends on the first constant term. To describe this structure, we follow
\cite{Jimbo:1981b,Jimbo:1982,Andreev:2000} and define the sectors of
the complex plane:
\begin{equation}
{\cal S}_j=\{z\in\mathbb{C}\, |\, (2j-5)\frac{\pi}{2} < \arg z <
(2j-1)\frac{\pi}{2} \}, 
\end{equation}
$j=1,2,...$. On each ${\cal S}_j$ we have the following asymptotic behavior for
the solutions of the system \eqref{eq:schlesinger}: 
\begin{equation}
\left.\Phi(z)\right|_j=G_j(z^{-1})\exp(\tfrac{1}{2}
z\sigma_3) z^{-\tfrac{1}{2}\theta_\infty\sigma_3},
\label{eq:infinity}
\end{equation} 
where $G_j(z^{-1})=\mathbbold{1}+{\cal O}(z^{-1})$ is analytic near
$z=\infty$. The Stokes phenomenon relates the solutions satisfying
\eqref{eq:infinity} between different sectors ${\cal S}_j$,
\begin{equation}
\Phi_{j+1}(z)=\Phi_j(z)S_j,
\end{equation}
where $S_k$ are the Stokes matrices. Now, $\Phi_{j}(e^{2\pi
  i}z)=\Phi_{j+2}(z)e^{-\pi i\theta_\infty\sigma_3}$ -- they are defined in the
same domain --, so we have that $S_{j+2}=e^{\pi
  i\theta_\infty\sigma_3}S_je^{\pi i\theta_\infty\sigma_3}$. Therefore one
can choose a basis where
\begin{equation}
S_{2j}=\begin{pmatrix}
1 & s_{2j} \\
0 & 1
\end{pmatrix},\quad
S_{2j+1}=\begin{pmatrix}
1 & 0 \\
s_{2j+1} & 1
\end{pmatrix},
\end{equation} 
where the numbers $s_k$ are called Stokes parameters. By the identification
of $S_{j+2}$ with $2\pi$ rotations, one then defines the monodromy at
infinity at sector ${\cal S}_j$ by
\begin{equation}
\left. M_\infty\right|_{{\cal S}_j}=S_jS_{j+1}e^{i\pi\theta_\infty\sigma_3},
\end{equation}
which is the one satisfying the free group relation with the other two
monodromy matrices:
\begin{equation}
M_\infty M_t M_0=\mathbbold{1}.
\end{equation}
Note that once we settle in a sector ${\cal S}_j$, say $j=1$, then
knowledge of only two consecutive Stokes parameters $s_1$ and $s_2$
are sufficient to reconstruct the whole series of Stokes matrices
$S_j$. 

The outcome of the above analysis is that the whole set of parameters
$\vec{\theta}=\{\theta_0,\theta_t,\theta_\infty,s_1,s_2\}$ is sufficient to determine
the monodromy matrices up to an overall conjugation. This set is thus called the
{\it monodromy data}.  

The existence of the monodromy matrices provides an explicit
representation of the 3-braid group -- in fact, the permutation group
$\mathbbold{S}_3$ -- acting on $M_i$ as
\begin{equation}
\begin{gathered}
\sigma_{ij}(M_i)=M_jM_iM_j^{-1}, \\
\sigma_{ij}(M_j)=(M_jM_i)M_j(M_jM_i)^{-1}.
\end{gathered}
\end{equation}
These generators satisfy
\begin{equation}
\sigma_{ij}\circ\sigma_{jk}\circ\sigma_{ki}=\sigma_{ik}\circ\sigma_{kj}
\circ\sigma_{ji},
\end{equation}
which is known as the Yang-Baxter relation \cite{Dubrovin2007a}. The
existence of the monodromy matrices then assures that the Rabi model
is integrable in the algebraic sense.

\section{The Isomonodromy Method}

The Riemann-Hilbert problem consists in finding a Fuchsian system with a
prescribed set of monodromies. Our problem is the inverse one. In order to
solve such inverse Riemann-Hilbert problem, we will first notice that there
are many different families of $A_i$'s which give the same
monodromy. Some of them are trivially related by overall
conjugation. But even so there is still a family of non-trivial set of
Fuchsian systems parametrized by the position of an extra singular
point $t$. 

This family was first described by Schlesinger -- see
\cite{Jimbo1981b,Jimbo:1981b,Jimbo:1981-3,Iwasaki:1991} for reviews --
but it is more easily understood in terms of flat holomorphic connections
\cite{Atiyah:1982}. Suppose we set 
\begin{equation}
\label{eq.schlesingersystemgen}
A(z,t)=\frac{1}{2}\sigma_3+\frac{A_0(t)}{z}+\frac{A_t(t)}{z-t}=\frac{\partial
  \Phi(z,t)}{\partial t}\Phi^{-1}(z,t)
\end{equation}
as the ``$z$-component'' of a flat connection. It is straightforward
to see that, if we consider the ``$t$-component'' as
\begin{equation}
B(z,t)=-\frac{A_t(t)}{z-t},
\end{equation}
then $F=\partial_tA-\partial_zB+[A,B]=0$ if the $A_i(t)$ satisfy
\begin{equation}
\begin{gathered}
\frac{\partial A_0}{\partial t}=\frac{1}{t}[A_t,A_0],\\[0.1cm]
\frac{\partial A_t}{\partial t}=-\frac{1}{t}[A_t,A_0]-\frac{1}{2}[A_t,\sigma_3].
\label{eq:schlesingerheun}
\end{gathered}
\end{equation}
This system are called the {\it Schlesinger equations}. Since the ``field
strenght'' $F$ vanishes, then the monodromy data of the Fuchsian
system \eqref{eq:schlesinger} will be independent of $t$ if $A_0(t)$
and $A_t(t)$ satisfy the equations \eqref{eq:schlesingerheun}. The matrices $A_0$ and $A_t$ can be
thought of as a Lax pair for the isomonodromy flow.

Despite being seemingly more complicated, the Schlesinger equations 
\eqref{eq:schlesingerheun} have a Hamiltonian structure. The most
direct way to illustrate it is to consider the EDO associated with the
generic Fuchsian system \eqref{eq.schlesingersystemgen}. Let us choose
a gauge such that
\begin{equation}
\Tr A_\infty = \theta_\infty.
\end{equation}
Consider the off-diagonal term $A_{12}$ of
\eqref{eq.schlesingersystemgen}. It is of the form 
\begin{equation}
A_{12}(z)=\frac{k(z-\lambda)}{z(z-t)},
\end{equation}
where $k,\lambda$ are linear functions of $(A_0)_{12}$ and
$(A_t)_{12}$. Now, by writing the solution as 
\begin{equation}
\Phi(z)=
\begin{pmatrix}
f^{(1)}_+(z) & f^{(2)}_+(z) \\
f^{(2)}_-(z) & f^{(2)}_-(z) 
\end{pmatrix},
\end{equation}
one can check that the elements of the first row $f^{(1,2)}_+(z)$ satisfy
\begin{equation}
\begin{gathered}
\frac{d^2}{dz^2}f^{(1,2)}_++p(z)\frac{d}{dz}f^{(1,2)}_++q(z)f^{(1,2)}_+=0, \\[5pt]
p(z)=\frac{1-\theta_0}{z}+\frac{1-\theta_t}{z-t}-\frac{1}{z-\lambda},
\\[5pt]
q(z)=-\frac{1}{4}+\frac{C_0}{z}+\frac{C_t}{z-t}+\frac{\mu}{z-\lambda},
\label{eq:edofromschlesinger}
\end{gathered}
\end{equation}
where $C_0$, $C_t$, $\lambda$ and $\mu$ are complicated functions of
the entries of $A(z)$. This EDO has, along with the singular points at
$z=0,t,\infty$, an extra singularity at $z=\lambda$. This singularity can be
checked to be an apparent one: the solutions of the indicial equation
at $z=\lambda$ gives $\alpha_\lambda^+=0,2$ and there is no logarithm
behavior due to the algebraic relation between the parameters:
\begin{equation}
\mu^2-\left[\frac{\theta_0-1}{\lambda}+\frac{\theta_t-1}{\lambda
    -t} \right]\mu+\frac{C_0}{\lambda}+\frac{C_t}{\lambda-t}=\frac{1}{4}.
\end{equation}
This relation means that the change of $t$ has to be accompanied by a
change in $\lambda$ and $\mu$ so that the relation is maintained. 

The Schlesinger system \eqref{eq:schlesingerheun}, when written
in these parameters, yield the Painlevé V equation for the following
function of the entries of the $A_i$: 
\begin{equation*}
y(t)=\frac{(A_0)_{11}(A_t)_{12}}{(A_t)_{11}(A_0)_{12}}= 
\frac{\theta_0+\theta_t-\theta_\infty-(2\mu-1)(\lambda-t)}{
  \theta_0+\theta_t-\theta_\infty-(2\mu-1)\lambda}.  
\end{equation*}
The Painlevé V is part of the family of second order differential
equations with rational coefficients and the Painlevé property: all the
branch points of the solutions are fixed, determined by the ODE
itself \cite{Miwa:1981,Iwasaki:1991}. These equations define new special
functions, and the Painlevé V in particular has been useful to compute
correlation functions of strongly coupled Bosonic systems
\cite{Jimbo:1980}, distribution functions of random matrix theory,
certain limits of conformal blocks and the $XY$ model -- see
\cite{Tracy:2011} for a (not exaustive) list of applications. It has
also been shown to give exact analytic expressions for the scattering
of massless fields in black hole backgrounds
\cite{Novaes2014,daCunha:2015ana}.

We follow \cite{Andreev:2000} and define the
tau-function:  
\begin{equation}
\frac{d}{dt}\log\tau(t,\vec{\theta}) = 
-\frac{1}{2}\Tr\sigma_3A_t -\frac{1}{t}\Tr A_0A_t 
\label{eq:ourtaufunction}
\end{equation}
which satisfies a third order non-linear ODE -- the so-called
$\sigma$-form of the Painlevé equations and it is defined up to a
multiplicative constant. 

The tau-function has the direct interpretation of
generating function for correlations in field-theoretic applications
of the Painlevé transcendents. Asymptotic expressions for the
tau-function have been derived in \cite{Jimbo:1982} and the
(irregular) conformal block interpretation was given in
\cite{Gamayun:2012ma,Gamayun:2013auu}, and the relevant results are
listed in the Appendix. In order to describe it we define the
composite monodromy parameter 
\begin{equation}
\begin{aligned}
2\cos\pi\sigma & =\Tr(M_tM_0) =\Tr(M_\infty^{-1}) \\
& =2\cos\pi\theta_\infty+ e^{\pi i\theta_\infty}s_1s_2,
\label{eq:compositemonodromy}
\end{aligned}
\end{equation}
then the monodromy data can be written as
$\vec{\theta}=\{\theta_0,\theta_t,\theta_\infty,\sigma,s_i\}$. We will
assume generic ({\it i.e.}, non-multiples of $\pi$) values for the
monodromy data $\vec{\theta}$ so these expressions can be locally
inverted. 

In terms of \eqref{eq:ourtaufunction}, the existence of monodromy
matrices for the Rabi Fuchsian system \eqref{eq:schlesinger} amounts
to the existence of a solution to the tau-function given the
initial conditions:
\begin{equation}
\begin{gathered}
\left.\frac{d}{dt}\log\tau(t,\vec{\theta})\right|_{t=-4g^2}
=\frac{E+g^2}{2}+ \frac{\Delta^2}{4g^2} \\
\left.\frac{d^2}{dt^2}\log\tau(t,\vec{\theta})\right|_{t=-4g^2}= 
\frac{1}{t^2}\Tr A_0A_t=\frac{\Delta^2}{16g^4} ,
\label{eq:solution}
\end{gathered}
\end{equation}
which is guaranteed on general grounds. Note that in the case of
interest $\theta_0=\theta_t=E+g^2$ and $\theta_\infty=0$, these
conditions can be inverted to yield the non-trivial monodromy data --
the Stokes parameters in our application. It should be stressed that
the equations above solve the Rabi model in an implicit but
combinatorial sense: the formulae given in the Appendix give an
asymptotic expansion for the Painlevé V tau-function near $t=0$. The
same special type of Painlevé V system as above was studied in a
series of papers \cite{McCoy1985a,McCoy1986a,McCoy1986}, where the
invariants of the isomonodromy flow were calculated, and, in the last
of the series, a Toda chain structure outlined.  

\section{Quantization}

The pair of equations \eqref{eq:solution} provides an implicit solution
of the Rabi model in terms of the Painlevé V tau-function. The
quantization condition for the energy can be solved in a similar form
by making use of the composite monodromy parameter
\eqref{eq:compositemonodromy}. We start by considering the asymptotics
of the solution of the Fuchsian system \eqref{eq:schlesinger} with the
behavior near $z=\infty$ given by \eqref{eq:infinity}. Near the
regular singular points $z=0,t$ the solution will behave as:
\begin{equation*}
\Phi(z)=
\begin{cases}
G_0z^{\left(\begin{smallmatrix}
\theta_0 & 0 \\
0 & 0 
\end{smallmatrix}\right)}(\mathbbold{1}+{\cal O}(z))C_0, \quad
z\rightarrow 0 \\[0.2cm] G_t(z-t)^{\left(\begin{smallmatrix}
\theta_t & 0 \\
0 & 0 
\end{smallmatrix}\right)}(\mathbbold{1}+{\cal O}(z-t))C_t, \quad
z\rightarrow t, 
\end{cases}
\end{equation*}
where $G_i$ ($i=0,t$) are the matrices diagonalizing $A_i$, and
$C_i$ are called the connection matrices. The monodromy matrices are 
diagonalized by the connection matrices, that is,
\begin{equation}
M_i=C^{-1}_ie^{\pi i\theta_i}C_i.
\end{equation}
Thus, the $C_i$ are defined up to right multiplication. Without loss
of generality we can take $\det C_i=1$. The matrix
$C_0C_t^{-1}$ can be seen to ``connect'' the natural solutions of the system
\eqref{eq:schlesinger} at $z=0$ and $z=t$.

We can use the monodromy matrix to solve for the eigenvalue
problem. The solution of \eqref{eq:schlesinger} is required from physical
grounds to be analytic on the whole plane -- it will have an essential
singularity at $z=\infty$, but analyticity at $z=0,t$ ensures that
the quantum state defined by the solution has finite expectation
values for the relevant physical quantities (like the bosonic number
operator). This condition is translated to our language by requiring
that the matrix $C_0C_t^{-1}$ which connects the natural solutions at
$z=0$ and $z=t$ is diagonal: the analytic solution at $z=0$ will also
be analytic at $z=t$. In principle the connection could be ``upper
triangular'': the second solution at $z=0$, which diverges as
$z^{\theta_0}$, could be connected to a superposition of the divergent
and the regular solutions at $z=t$, but one can easily see that this
does not happen: consider the determinant of the fundamental matrix,
$\det\Phi$, which satisfies the equation
\begin{equation}
\frac{d}{dz}\det\Phi =
\left(\frac{\theta_0}{z}+\frac{\theta_t}{z-t}\right)\det\Phi.
\end{equation}
This equation yields $\det\Phi=z^{\theta_0}(z-t)^{\theta_t}$, and can
be used to ``normalize'' the solutions, in the sense that now the
fundamental matrix has unit determinant. One can convince oneself that
this normalization does not change the connection matrices $C_i$, but
now the two natural solutions at any particular singular points have a
similar behavior: $(z-z_i)^{\pm\theta_i/2}$. This parity, a
$\mathds{Z}_2$ parity, was fundamental in Braak's work. In \cite{daCunha:2015ana}, it was though associated to a time-reversal symmetry. For our
application, this symmetry guarantees that the vanishing of one of the
off-diagonal elements of $C_0C_t^{-1}$ will imply the vanishing of the
other off-diagonal term. Hence $C_0C_t^{-1}$ will be diagonal.  

Now, a diagonal $C_0C_t^{-1}$ implies, for the composite
monodromy parameter $\sigma$, defined in
\eqref{eq:compositemonodromy}, that 
\begin{equation}
\cos\pi\sigma = \cos\pi(\theta_0+\theta_t).
\end{equation}
Therefore, the regularity of the solution is expressed as a
quantization condition: 
\begin{equation}
\label{eq:compositemonodromyfinal}
\sigma_n = 2n+\theta_0+\theta_t =2(E+g^2+n),\quad\quad
n\in\mathds{Z}. 
\end{equation}
Since $\sigma$ is given in terms of the Stokes parameters $s_{1,2}$,
this condition can be fed into the solution \eqref{eq:solution} to
yield the quantized values for the energy $E_n$. The completion of
this task requires the knowledge of the expansion of the
tau-function given in the Appendix. 

\section{Perspectives}

The methods described here are useful not only to show the existence
of the monodromy matrices, and hence Yang-Baxter integrability, for
the Rabi model but it also provides a solution for the eigenvalue
problem in terms of the transcendental equation \eqref{eq:solution}. 
Given that there is a combinatorial expansion of the Painlevé V
tau-function in terms of irregular conformal blocks, one can then
implement a numerical/symbolic computation to complete the task of
finding the eigenvalues, using the expansion given in the Appendix and
the quantization condition \eqref{eq:compositemonodromyfinal}.

Another interesting direction would be to use the proposed formalism
to other similar systems. For instance, the extension to the model
with broken parity introduced by Braak. This consists in adding a term
of the form $\gamma \sigma^x$ to the Rabi Hamiltonian, which seems to
be a simple extension and amenable through the methods described here. 

\section*{Acknowledgements}

The authors would like to thank Andrés Reyes-Lega for discussions and
the hospitality of Universidad de los Andes in Bogotá, Colombia, where
most of this work was conducted. BCdC thanks Alessandro Villar for
discussions and acknowledges partial support from PROPESQ/UFPE and
FACEPE under grant APQ-0051-1.05/15. 

\appendix

\section{Formulae for the Painlevé V $\tau$-function} 

Here we lift the relevant formulae from \cite{Gamayun:2013auu}. In the
following we consider the tau-function as defined in
\cite{Jimbo:1982}:
\begin{equation}
\tau(t)=t^{((\theta_0-\theta_t)^2-\theta_\infty^2)/4}[\tilde{\tau}(t)]^{-1}. 
\end{equation}
The expansion for the tau-function is of the form: 
\begin{equation}
\tilde{\tau}(t,\vec{\theta})=\\
\sum_{n\in\mathbb{Z}}C(\{\theta_i\},\sigma+n)s^nt^{(\sigma+n)^2}
{\cal B}(\{\theta_i\},\sigma+n;t),
\label{eq:tau5expansion}
\end{equation}
where the irregular conformal block ${\cal B}$ is given as a power
series over the set of Young tableaux $\mathbbold{Y}$:
\begin{equation}
{\cal
  B}(\{\theta_i\},\sigma;t)=e^{-\theta_tt}\sum_{\lambda,\mu\in\mathbbold{Y}} 
{\cal B}_{\lambda,\mu}(\{\theta_i\},\sigma)t^{|\lambda|+|\mu|}, 
\end{equation}
\begin{multline}
{\cal
  B}_{\lambda,\mu}=\prod_{(i,j)\in\lambda}\frac{(\theta_\infty+\sigma+i-j)
  ((\theta_t+\sigma+i-j)^2-\theta_0^2)}{h_\lambda^2(i,j)(\lambda'_j+
  \mu_i-i-j+1+2\sigma)} \times \\
\prod_{(i,j)\in\mu}\frac{(\theta_\infty-\sigma+i-j)
  ((\theta_t-\sigma+i-j)^2-\theta_0^2)}{h_\mu^2(i,j)(\lambda_i+
  \mu'_j-i-j+1+2\sigma)}.  
\end{multline} 
where $\lambda$ denotes a Young tableau, $\lambda_i$ is the number of
boxes in row $i$, $\lambda'_j$ is the number of boxes in column $j$
and $h_\lambda(i,j)=\lambda_i+\lambda'_j-i-j+1$ is the hook length
related to the box $(i,j)\in\lambda$. The structure constants $C$
are rational products of Barnes functions:
\begin{multline}
C(\{\theta_i\},\sigma)=\prod_{\epsilon=\pm}
G(1+\theta_\infty+\epsilon\sigma)
G(1+\theta_t+\theta_0+\epsilon\sigma) \times \\
G(1+\theta_t-\theta_0+\epsilon\sigma) /
G(1+2\epsilon\sigma),
\end{multline}
where $G(z)$ is defined by the functional equation
$G(1+z)=\Gamma(z)G(z)$. The parameters $\sigma$ and $s$ in
\eqref{eq:tau5expansion} are related to the ``constants of
integration'' of the Painlevé V equation. The $\sigma$ is the same
monodromy parameter as \eqref{eq:compositemonodromy}, whereas
$s$ has a rather lengthy expression in terms of monodromy data that
can be read from \cite{Jimbo:1982}. The Painlevé V tau-function was
also considered in great detail in \cite{Andreev:2000}. The particular
set of parameters considered here were also considered in
\cite{McCoy1985a,McCoy1986a,McCoy1986}.


%

\end{document}